\documentclass{aa}
\usepackage[varg]{txfonts}
\usepackage{graphicx}
\usepackage{xcolor}
\bibpunct{(}{)}{;}{a}{}{,} 

\begin{document} 

\title{ Experimental characterization of modal noise in multimode fibers for astronomical spectrometers.}
\titlerunning{ Experimental characterization of modal noise in fibers. }
\authorrunning{E. Oliva, M. Rainer, A. Tozzi et al.}

   \author{
           E. Oliva\inst{1}
          \and M. Rainer\inst{1}
          \and A. Tozzi\inst{1}
          \and N. Sanna \inst{1}
          \and M. Iuzzolino\inst{2}
          \and A. Brucalassi\inst{1}
          }
   \institute{
              INAF - Osservatorio Astrofisico di Arcetri,
              Largo E. Fermi 5, I-50125 Firenze, Italy
         \and
              Officina Stellare S.p.A.,
              via delle Tecnica 8789, I-36030, Sarcedo (VI), Italy
             }

   \date{Received September 02, 2019; accepted October 10, 2019}

 
  \abstract
{High resolution spectroscopy at high S/N ratios
is one the key techniques of the quantitative study of the atmospheres of extrasolar planets. 
Observations at near-infrared wavelengths with fiber-fed spectrographs coupled to extremely large telescopes are particularly important to tackle the ultimate goal of detecting biosignatures in rocky planets.
}
{
To achieve high S/N ratios in fiber-fed spectrogrpahs, the systematic noise effects introduced by the fibers must be properly understood and mitigated.
In this paper we concentrate on the effects of modal noise in multimode fibers.
}
{
Starting from our puzzling on-sky experience with the GIANO-TNG spectrometer we set up an infrared high resolution spectrometer in our laboratory and used this instrument to characterize the modal noise generated in fibers of different types (circular and octagonal) and sizes.
Our experiment includes two conventional scrambling systems for fibers: a mechanical agitator and an optical double scrambler.
}
{
We find that the strength of the modal noise primarily depends on how the fiber is illuminated. It dramatically increases when the fiber is under-illuminated, either in the near field or in the far field.
The modal noise is similar in circular and octagonal fibers. The Fourier spectrum of the noise decreases exponentially with frequency; i.e.,  the modal noise is not white but favors broad spectral features.
Using the optical double scrambler has no effect on modal noise. The mechanical agitator has effects that vary between different types of fibers and input illuminations. In some cases
this agitator has virtually no effect. In other cases, it mitigates the modal noise, but flattens the noise spectrum in Fourier space; i.e., the mechanical agitator preferentially filters the broad spectral features.
}
{
Our results show that modal noise is frustratingly insensitive to the use of octagonal fibers and optical double scramblers; i.e., the conventional systems used to improve the performances of spectrographs fed via unevenly illuminated fibers.
Fiber agitation may help in some cases, but its effect has to be verified on a case-by-case basis.
More generally, our results indicate that the design of the fiber link feeding a spectrograph should be coupled with laboratory measurements that reproduce, as closely as possible, the conditions expected at the telescope.
}

 \keywords{Astronomical instrumentation, methods and techniques -- Instrumentation: spectrographs --
   Techniques:miscellaneous -- Techniques:fibers --}

\maketitle
%

\section{Introduction}
Quantitative spectroscopy of the atmospheres of extrasolar planets is one of the key science cases of modern astrophysics.
High resolution spectrographs coupled to extremely large telescopes are particularly important to achieve the ultimate goal of detecting biosignatures in the atmospheres of earth-like planets  \citep[e.g.,][]{Maiolino_HIRES}. There are two main techniques:
transit spectroscopy, which is based on a time series of high resolution stellar spectra taken during the transit of a planet  \citep[e.g.,][]{Brogi2012}; and direct light detection, which is based on simultaneous high resolution spectra taken through an integral field unit (IFU) that includes the star and surrounding exoplanet(s). The contrast between the planet and the stellar signals can be boosted using adaptive optics correction \citep[e.g.,][]{Snellen2015}.
In either case the spectrum of the exoplanet atmosphere consists of a large number of narrow ($\simeq$ few km/s) lines with amplitudes orders of magnitude weaker than the signal from the star.
For earth-like planets orbiting M dwarfs the amplitude of the planet spectrum is only a few $\times 10^{-5}$ of the stellar continuum.
Cross-correlation techniques are used to identify the exoplanet spectrum taking advantage of the Doppler shift introduced by the orbital motion of the planet.\\
%

   \begin{figure*}
   \includegraphics[width=\hsize]{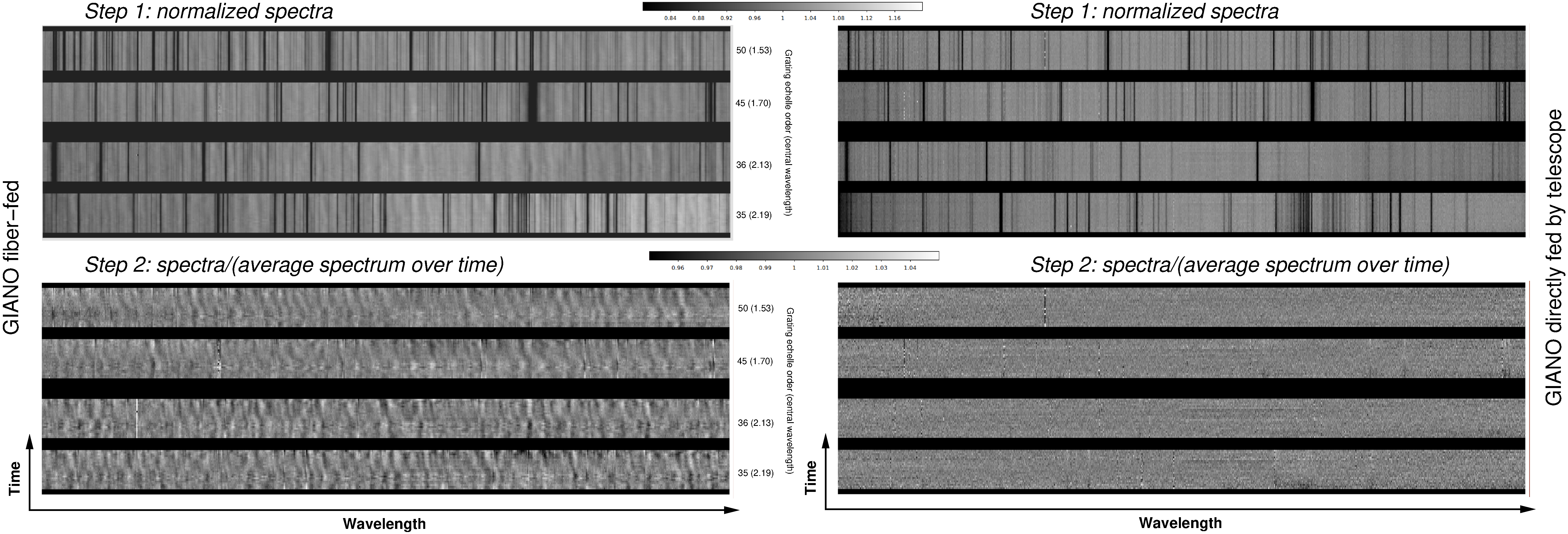}
      \caption{
         Times-series spectra of stars collected with the GIANO spectrometer.
         The left- and right-hand panels show the data collected with and  without fiber feeding, respectively. Modal noise causes the wavy pattern clearly visible in the data taken with fibers.
              }
         \label{fig_giano_data}
   \end{figure*}
%

These observations are normally simulated assuming that the spectral data are only affected by detector-related noises (readout and dark-current noise) and photon noise. 
Recently, \citet{Hawker2019} investigated the effects systematic extra noise introduced by a non-perfect correction of the telluric features.
Other systematic effects introduced by the instrument are evaluated based on practical experience with existing spectrometers.
No detailed attempt is made to identify the technical aspects of the instruments that are critical, useful, or just redundant for this type of measurements.

In this paper we concentrate on the effects of modal noise in spectrographs fed via multimode fibers.
This phenomenon is related to the interference between the light modes that propagate through a fiber. 
Inside a spectrometer, the interference pattern translates into a non-flat, wavy spectral distribution. 
Noise is generated whenever the interference pattern changes between different exposure; i.e., when the wavy pattern cannot be corrected using flat-field calibrations.

From the theoretical point of view the available literature on modal noise can be classified into two main groups.
The first group does not consider how the fiber-input is illuminated.
For example, \citet{Grupp2003}, \citet{Chen2006}, and \citet{Lemke2011} modeled the effect of modal noise introduced by vignetting occurring in the spectrometer, downstream of the fiber.
These authors unanimously concluded that the effect can be averaged
out and effectively mitigated by continuously agitating or stretching the fiber.
The second group mainly concentrates on how the fiber is illuminated. The most remarkable example is that of \citet{corbett2016} 
who argued that the abovementioned theories must be called into question because of the changes in the statistics of modal noise with arbitrary illumination of the fiber. These authors also mentioned that fiber agitation may have different effects, depending on how the fiber is illuminated. 
Other instructive examples are those of \citet{Mahadevan2014} and \citet{Halverson2014} who studied the best strategies to illuminate the fibers from the calibration sources, while \citet{Blind2017} studied the effects of a nonuniform illumination of the fiber carrying the stellar light.\\

From the experimental point of view there are many papers that concentrate on the importance of mechanical agitation of the fiber, from the earliest work of \citet{Baudrand2001} to the comprehensive characterization of different electromechanical devices of \citet{Petersburg2018}.\\
Mechanical agitators are often included in the fiber links for the calibration lights; while they are only rarely used for the fibers carrying the stellar light from the telescope. 
The spectrographs HARPS and ESPRESSO of the European Southern Observatory (ESO) 
are the most remarkable examples of spectrometers that do not shake the science fibers.
A similar approach is also foreseen for the 
Near Infrared Planet Searcher (NIRPS) under construction for the ESO 3.6 m telescope 
\cite[see][]{Blind2017}.
Other types of ``static scramblers'' (i.e., octagonal fibers and optical double scramblers) are favored in these instruments, even though their effects on modal noise are virtually unknown. Curiously enough, \citet{Mahadevan2014} argued that these static devices have no useful effect on modal noise generated in fibers carrying coherent light from  calibration sources.\\

 In this paper we address the  questions of under which circumstances modal noise is important, how modal noise
affects observations that require very high S/N ratios, and which technical solutions are useful to mitigate the effect of modal noise.
We start describing in Sect. \ref{section_giano} our puzzling experience with data collected with the infrared spectrometer GIANO \citep{giano14} at the Telescopio Nazionale Galileo (TNG) of the Roque de Los Muchachos Observatory in La Palma.
We then describe the laboratory spectrometer that we specifically built to study the phenomenon (Sect. \ref{section_labspec}). The results are presented in Sect. \ref{section_results}. In Sect. \ref{section_discussion} we discuss the results and draw our conclusions.

\section{Modal noise in GIANO spectra with fibers}
\label{section_giano}

The spectrograph GIANO is the high resolution (R$\simeq$50,000) infrared (950--2450 nm) spectrometer of the TNG telescope \citep{giano14}. The GIANO instrument was designed and built for direct feeding of light at a dedicated focal station.
In 2012, after a reorganization of the focal stations at the TNG, the instrument was provisionally commissioned and used in the GIANO-A configuration: with the spectrometer positioned on the rotating building and fed via a pair of $\oslash$~85~$\mu$m fibers connected to the only available focal station \citep{tozzi14}.
In 2016 the spectrometer was eventually moved to the originally foreseen configuration (called GIANO-B), in which the instrument is directly fed from the telescope (i.e., no fibers). This configuration can also be used for simultaneous observations with HARPS-N: the so-called GIARPS mode \citep{tozzi16}.

The fiber link used in the provisional GIANO-A configuration could not be fully optimized for astronomical observations. One of the main limitations was modal noise. This was already evident in the very first acquisitions of calibration frames: flats taken at different times showed variable, wavy spectral features. 
To cure this problem we introduced a mechanical agitator in the fiber link carrying the light to the spectrometer. This solved the problem of the calibration light: comparisons of various groups of flat exposures always yielded noise values compatible with the theoretical photon noise, even for S/N ratios exceeding 3000.
However, much to our surprise, the mechanical agitator did not improve the quality of the stellar spectra, even though the scientific and calibration lights went through the same fibers and mechanical agitator.
All our attempts to improve the situation by modifying the agitation mechanisms were unsuccessful.

The effect of modal noise is best visible in data aimed at detecting exoplanet features during transits. The observations consist of a continuous series of spectra of the same star with the instrument in the same configuration \citep{Snellen2010}.
The spectra are organized in 2D frames with the wavelength on the X-axis and time on the Y-axis. 
The ``Step 1'' images are made by combining the normalized spectra: they include the lines intrinsic to the star (stable over time) and the telluric absorption features, whose depths vary with time and air mass.
Both families of lines are removed (to a first approximations) by a linear fit with air mass of each wavelength channel: this generates the ``Step 2'' frames.

The data shown in Fig.~\ref{fig_giano_data} are an instructive comparison of such spectral series acquired using the GIANO spectrometer with and without fibers. 
The irregular, wavy pattern in the data taken with fibers is the effect of modal noise. The peak-to-valley amplitude of these features is about 10\% of the continuum level.
In spite of this large noise, these data were successfully used by \citet{Brogi2018} to detect water in the HD189733b. Data with a similar pattern of modal noise were used to detect methane in HD102195b \citep{Guilluy2019}.
These results are encouraging because they demonstrate that the signal of the exoplanet atmosphere can be detected even in situations in which the modal noise strongly limits the final quality of the data. More specifically, these results indicate that modal noise primarily produces broad features, while it does not seem to affect the measurement of narrow (unresolved) lines such as those produced in the atmosphere of exoplanets.

   \begin{figure}
   \includegraphics[width=\hsize]{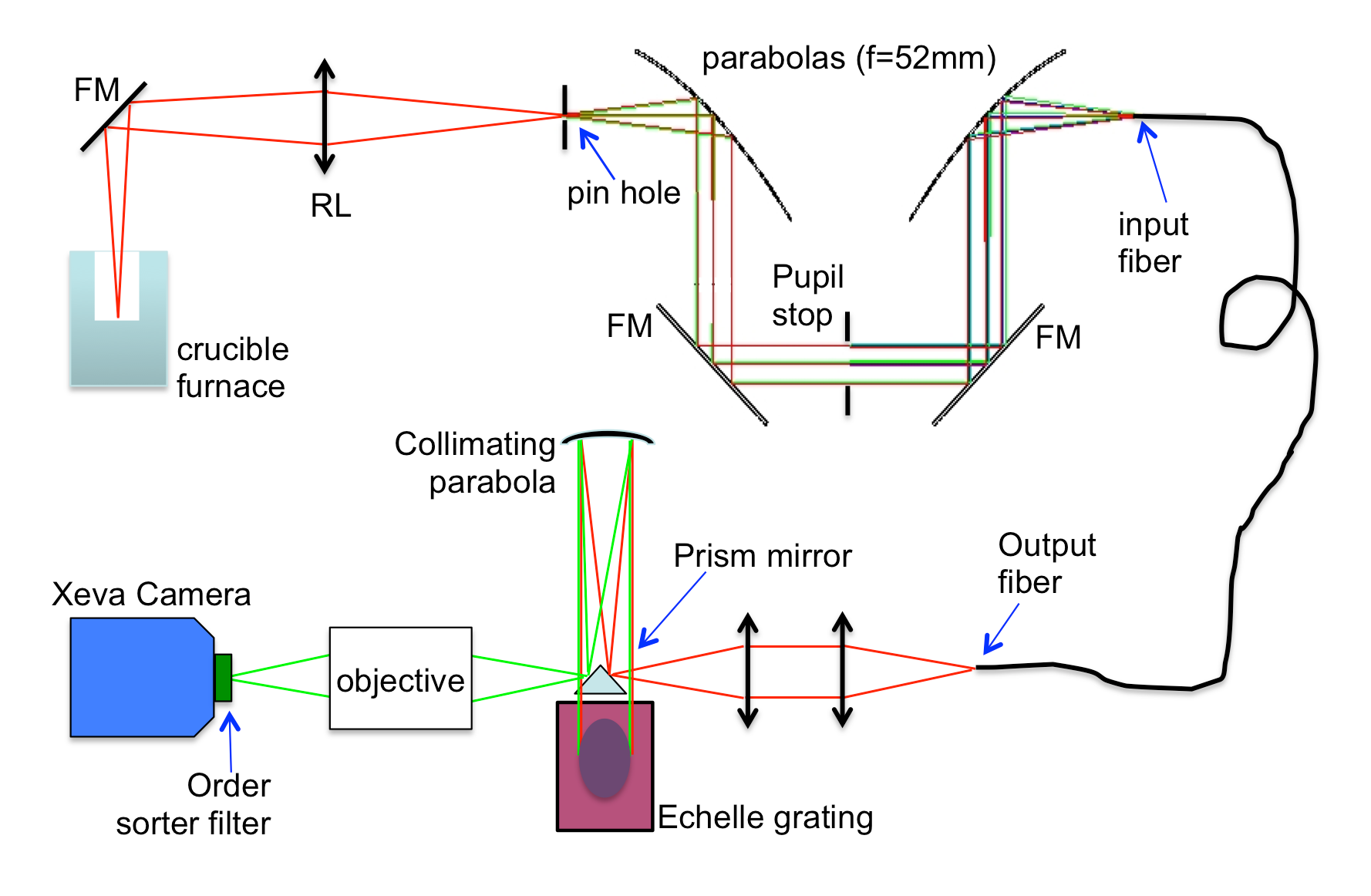}
      \caption{
          Schematic layout of the spectrometer used for the characterization
          of the modal noise.
              }
         \label{fig_spec_lab}
   \end{figure}

\section{Characterizing modal noise in the laboratory}
\label{section_labspec}

Given the puzzling results obtained at the telescope, we decided to investigate the origin and effects of modal noise using a dedicated spectrometer in the laboratory. 
The instrumental setup is summarized in Figure~\ref{fig_spec_lab}; see also \citet{Iuzzolino2014,Iuzzolino2015}. 
The light source is a commercial furnace at 1150~$^o$C. This somewhat inconvenient device is the only light source that guarantees a sufficiently bright and uniform illumination both in the near field and far field. 

The light is injected into the fiber via a 1:1 optical relay with a field of view of 2 mm and a focal aperture of F/2; i.e., adequate for a complete filling of all the fiber modes.
The near-field illumination of the fiber can be limited by inserting a field stop (pinhole) at the entrance, while the far-field illumination can be regulated using a pupil stop (iris) in the collimated beam. Both stops can be moved laterally to modify the illumination pattern of the fiber input.

The fiber exit feeds a high resolution spectrometer that uses the spare grating of GIANO: a R2 echelle with 23.2 lines/mm. The grating angle is adjusted to obtain a central wavelength of 1600~nm at order 48. The light in the other orders is blocked by a narrowband filter positioned in front of the detector. The detector is a commercial InGaAs camera with 320x256 pixels of 30 microns size. The spectral dispersion is 0.032~nm/pixel.

Optionally, the fiber can be directed through a mechanical agitator that vigorously shakes a 30~cm section of the fiber at about 20 Hz. 
The other optional subsystem is an optical double scrambler that couples the light between two identical fibers, reverting the near and far fields. This device was originally designed for the $\oslash$~85~microns fibers of GIANO-A.

The spectra were collected with three types of fibers as follows: first, fused silica fibers with circular core $\oslash$~85~$\mu$m (from the same batch used for the WEAVE instrument); second, fused silica fibers with octagonal core with 67~$\mu$m pitch (from the stock catalog of Ceramoptec); and third, fused silica fibers with circular core $\oslash$~50~$\mu$m (from the catalog of Thorlabs).
In all cases, the fibers are mounted in 8~m long patches with 
commercial SMA connectors. 
Two such patches were alternatively used for each type of fiber. The results obtained with the two patches are indistinguishable.

\section{Results from laboratory spectra}
\label{section_results}

The laboratory spectra were organized in continuous time series of 1200 exposures each with 0.5 s integration time. The spectra were analyzed using a technique similar to that employed for the transits of exoplanets. 
The average of the first 200 spectra of the series were used as reference flat. All the spectra of the series were divided by the reference flat and renormalized (i.e., divided by their median value). Therefore, they are equivalent to the Step 2 data described in Sect.~\ref{section_giano} and Figure~\ref{fig_giano_data}.

\begin{figure*}
   \includegraphics[width=\hsize]{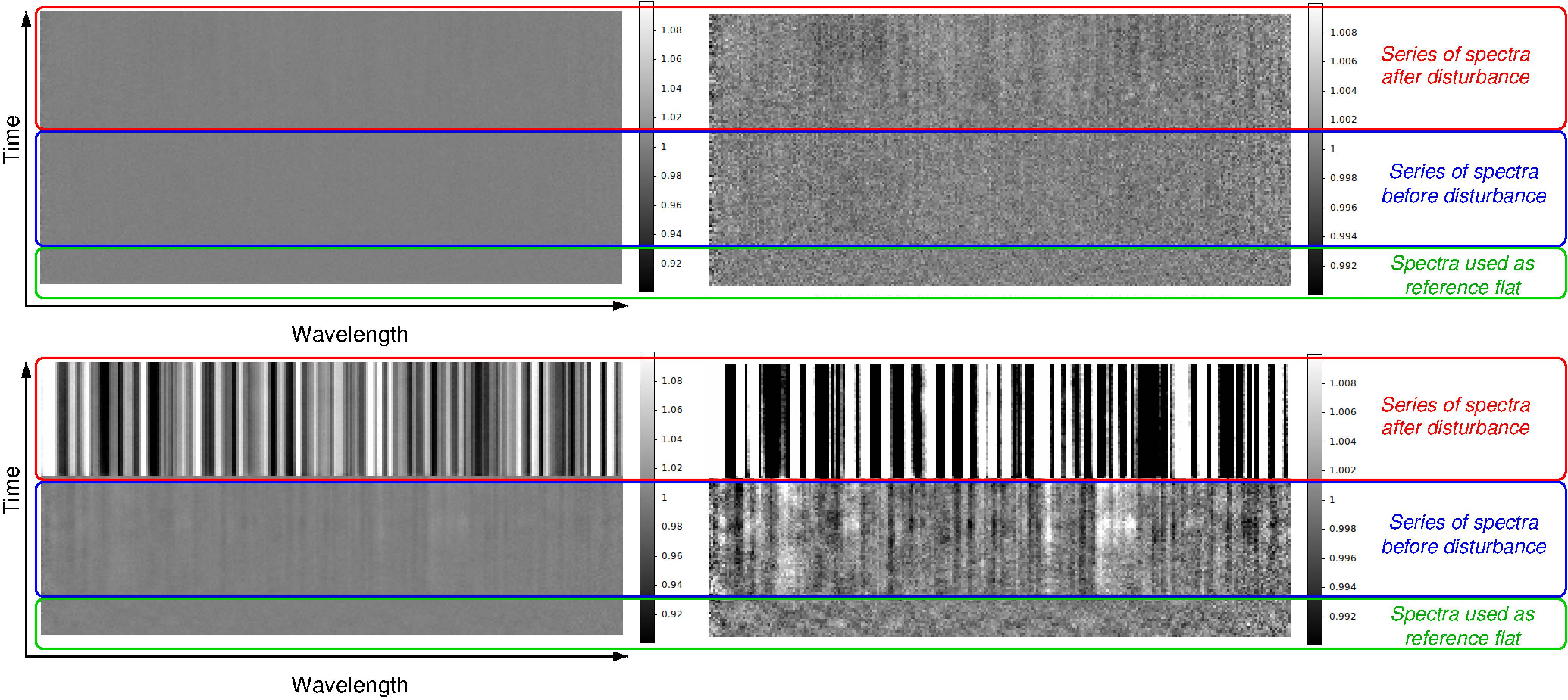}
      \caption{
         Example of 2D times-series spectra with different illuminations of the fiber. Top panel: uniform illumination (overfilled). Bottom panel: under-filled illumination. 
         The left- and right-hand panels show the same data with different gray scale limits: $\pm$10\% and $\pm$1\%, respectively. 
              }
         \label{fig_lab_spectra_2D}
   \end{figure*}
%

\begin{table*}[]
  \centering
  \begin{tabular}{|c|c|c|c|c|c|c|}
  \hline
    Fiber & Double scrambler & Shaker & Under-fill & Disturbance & $a$ & $s$ \\
  \hline
    Circular $\oslash$85~$\mu$m & no & no & near-field & changed illumination  & -2.0 & 5.9 \\
    Circular $\oslash$85~$\mu$m & no & no & near-field & moved fiber & -2.1 & 6.1 \\
    Circular $\oslash$85~$\mu$m & no & no & far-field & changed illumination  & -1.7 & 6.0 \\
    Circular $\oslash$85~$\mu$m & no & no & far-field & moved fiber  & -1.7 & 5.9 \\
    Circular $\oslash$85~$\mu$m & no & yes& near-field & changed illumination  & -3.6 & 4.0 \\
    Circular $\oslash$85~$\mu$m & no & yes& near-field & moved fiber & -4.0 & 3.9 \\
    Circular $\oslash$85~$\mu$m & no & yes& far-field & changed illumination  & -3.8 & 4.1 \\
    Circular $\oslash$85~$\mu$m & no & yes& far-field & moved fiber  & -4.1 & 3.9 \\
    Circular $\oslash$85~$\mu$m & yes & no& near-field & changed illumination  & -1.9 & 6.0 \\
    Circular $\oslash$85~$\mu$m & yes & no& near-field & moved fiber & -2.1 & 5.9 \\
    Circular $\oslash$85~$\mu$m & yes & no& far-field & changed illumination  & -1.6 & 6.1 \\
    Circular $\oslash$85~$\mu$m & yes & no& far-field & moved fiber  & -1.7 & 6.0 \\

    Circular $\oslash$50~$\mu$m & no & no & near-field & changed illumination  & -1.5 & 6.0 \\
    Circular $\oslash$50~$\mu$m & no & no & near-field & moved fiber & -2.0 & 5.8 \\
    Circular $\oslash$50~$\mu$m & no & no & far-field & changed illumination  & -1.4 & 6.1 \\
    Circular $\oslash$50~$\mu$m & no & no & far-field & moved fiber  & -1.8 & 5.9 \\
    Circular $\oslash$50~$\mu$m & no & yes& near-field & changed illumination  & -2.0 & 5.9 \\
    Circular $\oslash$50~$\mu$m & no & yes& near-field & moved fiber & -2.3 & 6.0 \\
    Circular $\oslash$50~$\mu$m & no & yes& far-field & changed illumination  & -1.7 & 5.8 \\
    Circular $\oslash$50~$\mu$m & no & yes& far-field & moved fiber  & -2.0 & 5.9 \\

     Octagonal 67~$\mu$m & no & no & near-field & changed illumination  & -2.1 & 6.1 \\
     Octagonal 67~$\mu$m & no & no & near-field & moved fiber & -2.0 & 5.8 \\
     Octagonal 67~$\mu$m & no & no & far-field & changed illumination  & -1.6 & 5.9 \\
     Octagonal 67~$\mu$m & no & no & far-field & moved fiber  & -1.5 & 6.0 \\
     Octagonal 67~$\mu$m & no & yes& near-field & changed illumination  & -2.4 & 4.0 \\
     Octagonal 67~$\mu$m & no & yes& near-field & moved fiber & -2.6 & 4.1 \\
     Octagonal 67~$\mu$m & no & yes& far-field & changed illumination  & -2.4 & 4.0 \\
     Octagonal 67~$\mu$m & no & yes& far-field & moved fiber  & -2.7 & 4.1 \\
\hline

    \end{tabular}
    \caption{Average values of strength ($a$) and slope ($s$, see Eq.~2) of the DFT spectrum of modal noise generated under different conditions. Typical scatters within each group of data are $\pm$0.3 for $a$ and $\pm$0.1 for $s$. See Sect.~\ref{section_results} for details}
    \label{tab_results}
\end{table*}

 \begin{figure*}
   \includegraphics[width=\hsize]{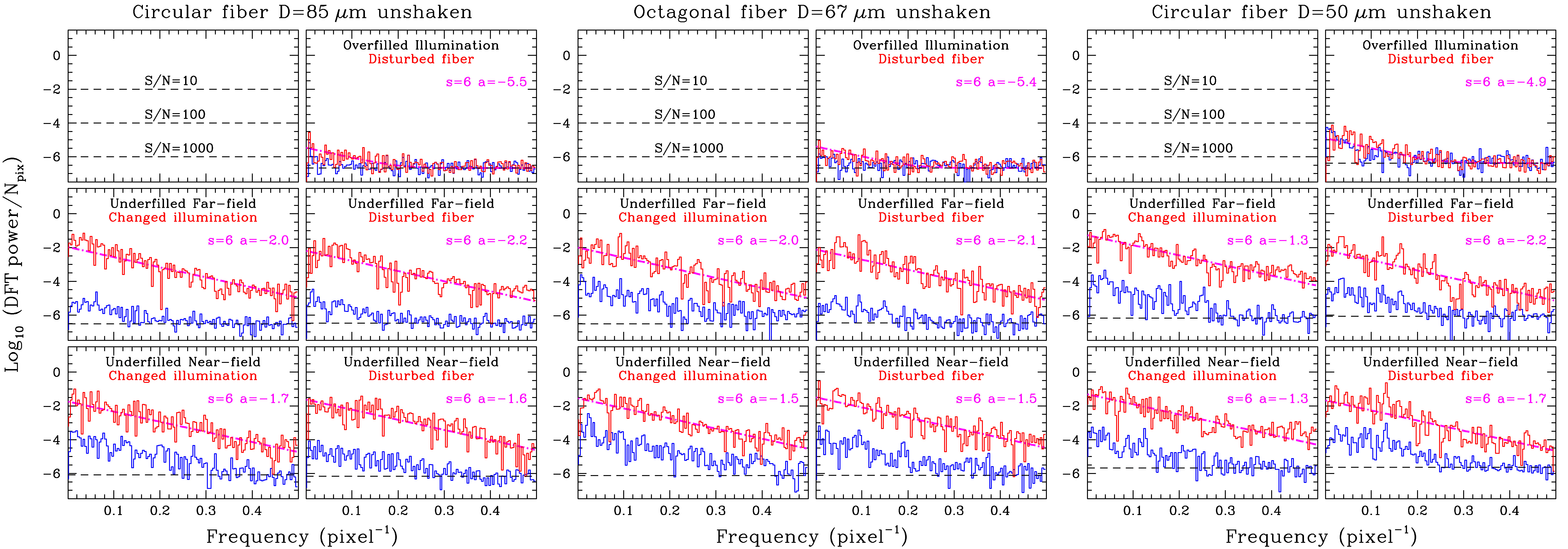}
      \caption{
          Digital Fourier transform spectral analysis of data taken with different fibers without mechanical agitator. See text (Sect.~\ref{section_results}) for details
              }
         \label{fig_FFT_01}
   \end{figure*}
%

\begin{figure*}
   \includegraphics[width=\hsize]{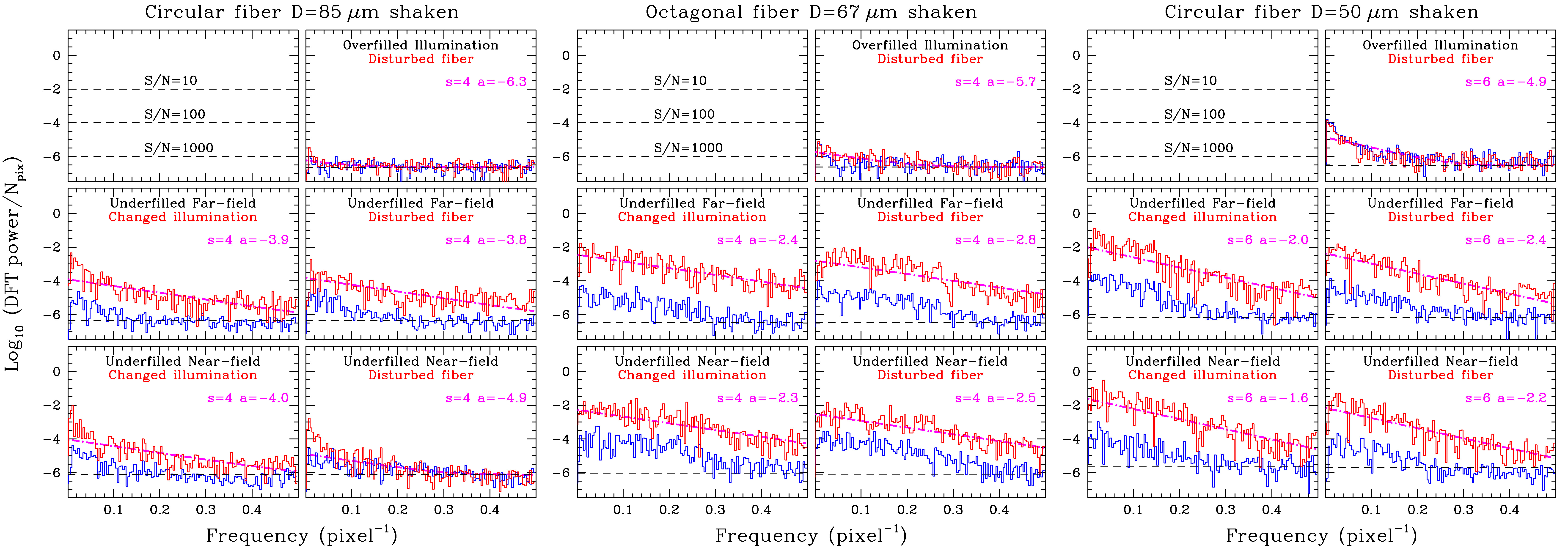}
      \caption{
          Same as Fig.~\ref{fig_FFT_01} for spectra with mechanical agitation of the fiber.
              }
         \label{fig_FFT_02}
   \end{figure*}
%

\begin{figure}
   \includegraphics[width=\hsize]{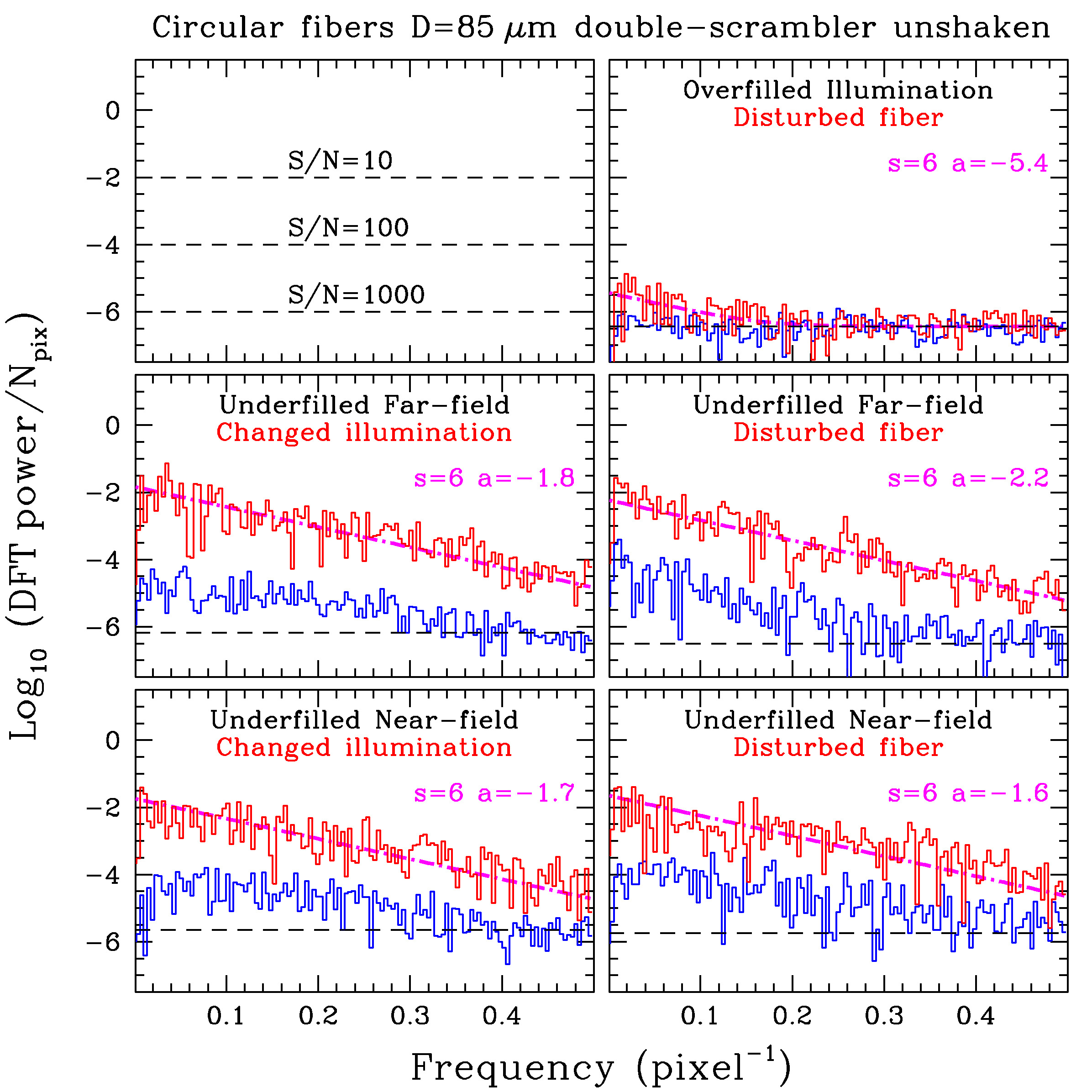}
      \caption{
          Same as Fig.~\ref{fig_FFT_01} for spectra with the optical double scrambler inserted.
              }
         \label{fig_FFT_03}
   \end{figure}
%

We took the first 700 spectra with the system in a undisturbed configuration, thereby avoiding in particular any movement of the fiber.
At this point, we intentionally generated a quick disturbance in the system. Therefore, the effect of disturbance was recorded in the remaining 500 spectra of the series. 
There are three types of disturbance as follows: 
\begin{enumerate}
\item Change of input illumination with the fiber under-illuminated in the near field. A pinhole of 10 microns is used as field stop. 
The disturbance consists of laterally moving the pinhole by 20~microns.
\item Change of input illumination with the fiber under-illuminated in the far-field. A pupil stop on the collimated beam limits the focal aperture to F/15. 
The disturbance consists of laterally moving the pupil stop by twice its diameter.  
\item Mechanical disturbance to the fiber. This consists of gently lifting and laying back a portion of the fiber on the optical bench. This disturbance is applied for any type of fiber illumination; i.e., under-illuminated and overfilled.
\end{enumerate}

Figure~\ref{fig_lab_spectra_2D} shows representative examples of such 2D spectra taken under different conditions.
The top spectrum is taken with overfilled illumination of the fiber. This spectrum is dominated by white noise and only shows a very weak pattern of extra noise after the disturbance. 
The bottom spectrum is taken with the fiber under-illuminated in the near field. This spectrum is characterized by a very prominent modal noise after the disturbance and also shows a weak pattern of modal noise in the spectra taken before the disturbance. Most probably, this is caused by tiny thermo-mechanical instabilities within the fiber.
%
%
In all spectra the modal noise seems to be dominated by low spatial frequencies, i.e., relatively broad features.\\

A quantitative analysis of the modal noise can be conveniently performed by computing the discrete Fourier transform (DFT) of the extracted 1D spectra
before and after the disturbance. Representative results are shown in Figures~\ref{fig_FFT_01}, \ref{fig_FFT_02}, and \ref{fig_FFT_03}. These plots show the variation of the logarithm of the average DFT power with spatial frequency (pixel$^{-1}$). In such a diagram, spectra dominated by white noise would appear as horizontal lines with constant value equal to the inverse square of the S/N ratio, as indicated in the top left insert of the figures.

We first concentrate on the spectra shown in the bottom left panel of Fig.~\ref{fig_FFT_01}. These spectra refer to measurements using a $\oslash$85~$\mu$m circular fiber without mechanical agitation. The fiber is under-illuminated in the near field and is disturbed by moving the input illumination.
The horizontal (black) dashed line at the bottom is the white noise level expected from the spectrometer alone.
This is the limiting noise of the measurement and is given by
$$ rms_{spectrometer} = \frac{\sqrt{2\, N_{frames}\, ( n_e + RON^2 + n_{dark} )}}{N_{frames}\  n_e} \eqno(1)  $$
where $n_e$ is the number of photoelectrons collected in each frame, $RON$ is the readout noise, $n_{dark}$ is the number of electrons produced by the detector dark current in each frame, and $N_{frames}$ are the number of exposures summed to obtain the 1D spectrum. The factor of 2 is included to take into account the noise introduced by dividing for the flat-field, which also consists of the sum of $N_{frames}$. We verified that the limiting noise of the spectra scaled as predicted by the above relationship down to values of $rms_{spectrometer}\simeq 10^{-4}$, equivalent to S/N ratios of $10^4$. These measurements were performed without fibers, i.e., with the spectrometer directly illuminated through a pinhole. \\
The bottom (blue) histogram is the measured DFT in the undisturbed part of the spectral series.
The top (red) histogram is the measured DFT in the disturbed part. The latter is fitted using an exponential function as follows:
%
$$ DFT_{total} = \left( DFT_{modal-noise}^{-2} + rms_{spectrometer}^{-4} \right)^{-1/2} \eqno(2)$$
$$ DFT_{modal-noise} = 10^a \cdot 10^{-s\cdot \nu} \eqno(3) $$
where $\nu$ is the frequency (pixel$^{-1}$). The fitted strength ($a$), its slope ($s$), and the function $DFT_{total}$ are displayed in magenta. 

The other panels of Fig.~\ref{fig_FFT_01} show the same type of data for various types of fibers, illumination and disturbances. The plotted spectra are randomly selected among a large set of data taken under similar conditions. The numerical results are report in Table~\ref{tab_results}.
All these data yield very similar results that can be summarized as follows.
\begin{itemize}
    \item Modal noise is very weak and barely measurable when the illumination of the fiber is overfilled both in the near field and far field (top right panels). Indeed, the ideal condition of overfilling is very difficult to achieve in practice; in our case it required using a furnace as light source (see Sect.\ref{section_labspec}).
    \item The strength of modal noise (parameter $a$ of Eq.~3) is similar in all the spectral series with under-illuminated fibers. The average value is $a =-1.8 \pm 0.3$ and is the same for all fibers types and measurement setups. In particular we do not find any significant difference between octagonal and circular fibers.
    \item The slope of the fit to the modal noise (parameter $s$ of Eq. 3) is remarkably similar in all the measurements. Its average value is $s=6.0\pm0.1$ without any significant difference between different types of fibers, illumination and disturbance.
\end{itemize}

Representative spectra obtained using the mechanical agitator of the fiber are shown in Fig.~\ref{fig_FFT_02} and the derived parameters are listed in Table~\ref{tab_results}. The results strongly depend on the type of fiber.
\begin{itemize}
    \item In the spectra taken with the $\oslash$85~$\mu$m fibers, the strength of the modal noise decreases by about two orders of magnitudes. 
    The slope of the DFT distribution becomes shallower with $s=4.0 \pm 0.1$.
    \item In the spectra taken with the 67~$\mu$m octagonal fibers, the strength of the modal noise decreases by only 0.7 dex and its slope decreases to $s=4.0 \pm 0.1$, i.e., the same flattening seen with the larger circular fiber. 
    \item In the spectra taken with the $\oslash$50~$\mu$m fibers, the average strength and the slope of the modal noise are similar to those observed without the mechanical agitator.
\end{itemize}
Noticeably, any decrease of modal noise is coupled with a flattening of its DFT noise spectrum. In other words, mechanical agitation (when it works) preferentially filters the broadest features of the modal noise.\\

Representative spectra obtained with an optical double scrambler are shown in Fig.~\ref{fig_FFT_03} and the derived parameters are listed in Table~\ref{tab_results}. 
The strength and slope of the noise spectrum are identical within the errors to the measurements without this device (Fig~\ref{fig_FFT_01}). This indicates that the double scrambler has no effect on the modal noise.

\section{Discussion and conclusions}
\label{section_discussion}
To the best of our knowledge, the data presented in this paper are the first systematic attempt to characterize modal noise in the laboratory using multimode fibers feeding a high resolution spectrometer similar to those designed (or built) for astronomical observations at large telescopes.
To emphasize the effects, the data are collected using unusually long wavelengths, relatively small fiber sizes, and somewhat extreme conditions for the input illumination of the fiber. 
More specifically, the number of propagating modes in our experiments vary between $\simeq$10 and 700, while most fiber-fed astronomical spectrographs work with several thousands modes.
The most important results are as follows. 

First, modal noise primarily depends on how the fiber is illuminated. This is in line to what postulated by \citet{corbett2016}.

Second, octagonal fibers and optical double scramblers do not seem to have any effect on modal noise. This is in line with what was already suggested by \citet{Mahadevan2014}. It implies that well-known technical solutions, employed with great success in current instruments optimized for accurate radial velocity measurements, are not necessarily useful in spectrographs that work in a parameter space in which modal noise is much more critical. 

Third, mechanical agitation of the fiber may help to decrease the strength of the modal noise. However, shaking has a remarkably different effects on fibers of different sizes and types. Moreover, when it works it flattens the DFT noise spectrum; i.e., it preferentially filters broad spectral features. Therefore, any trade-off study on mechanical agitators should include a quantitative analysis of the DFT noise spectrum. 

Fourth, modal noise can be decreased below the required S/N ratio by uniformly illuminating the fiber
both in the near field and far field. However, this is a very difficult condition to obtain in practice even for calibration sources \citep[see, e.g.,][]{Mahadevan2014}. For the science fibers coupled to the telescope we have to find a reasonable compromise between the total throughput and the modal noise generated by the various effects that contribute to under-illumination (e.g., telescope central obscuration, and point spread function variations).This task should ideally be performed in the laboratory using a realistic simulator of the telescope, fiber link, and spectrometer.

\begin{acknowledgements}
We thank the anonymous referee for helpful comments and suggestions.
We are grateful to Isabelle Guinouard for providing us with the $\oslash$85~$\mu$m fibers.
This research has been partly supported by INAF through the TECNO-INAF 2011 and ``FRONTIERA-2016'' competitive projects.
Based on observations made with the GIANO spectrometer at the Italian Telescopio Nazionale Galileo (TNG) operated on the island of La Palma by the Fundación Galileo Galilei of the INAF at the Spanish Observatorio del Roque de los Muchachos of the Instituto de Astrofisica de Canarias.
\end{acknowledgements}

\bibliographystyle{aa}
\bibliography{biblio}
\end{document}